\documentclass[12pt,a4paper]{article}

\usepackage[utf8]{inputenc}
\usepackage[T1]{fontenc}
\usepackage{amsmath,amssymb}
\usepackage{graphicx}
\usepackage{tikz}
\usetikzlibrary{patterns,decorations.pathreplacing}
\usepackage[margin=2.5cm]{geometry}
\usepackage{setspace}
\usepackage{float}
\usepackage{hyperref}
\title{Energy Poverty as a Structural Trap: \\The Role of Housing Efficiency and Non-Convex Technology}
\author{Nazaria Solferino, Universitas Mercatorum}
\date{}

\begin{document}
\maketitle

\begin{abstract}
Energy poverty persists even among households that are not income-poor, suggesting a deeper mechanism than mere budget constraints. We develop a model in which indoor thermal comfort is produced through a non-convex technology that couples energy input with dwelling efficiency. A critical efficiency threshold emerges below which the minimum comfort level is physically unattainable, regardless of how much energy is purchased. Households below this threshold suffer from \emph{structural} energy poverty, which income transfers alone cannot cure. The model yields three sharp policy predictions: energy price shocks are strongly regressive, efficiency investments dominate income transfers and price subsidies in reducing energy poverty, and a cost-effective anti-poverty strategy must combine targeted retrofits with temporary income support. The results are illustrated with symbolic diagrams and formal proofs.

\vspace{0.3cm}

\vspace{0.1cm}
\end{abstract}
\section{Introduction}
In the European Union, around 9\% of the population reported being unable to keep their home adequately warm in 2024, a figure that has remained stubbornly high despite modest economic growth \cite{Eurostat_Energy_Poverty_2024}. National observatories, such as the Italian OIPE, estimate that over 2 million households experience energy deprivation, with peaks above 12--18\% in Southern regions \cite{OIPE_2024}. These numbers coexist with a stable at-risk-of-poverty rate of around 16\% \cite{Eurostat_AROP_2025}, demonstrating that energy poverty is not a simple reflection of monetary poverty. The problem is exacerbated by large cross-country differences in energy prices, which in 2024 ranged from less than 80~EUR/MWh to over 110~EUR/MWh for electricity \cite{EC_energy_prices}.

The academic and policy literature has approached energy poverty from several directions. Early conceptual work by \cite{Boardman1991} defined fuel poverty as the need to spend more than 10\% of household income on energy to maintain a healthy indoor environment. \cite{Hills2012} refined the metric by introducing a “low income – high cost” indicator that remains influential in the UK. More recently, multidimensional frameworks have been developed by \cite{Bouzarovski2015}, who propose a global perspective on energy deprivation, and by \cite{Thomson2016}, who review definitions across the EU and highlight the need for a common operational concept. On the empirical side, \cite{Papada2016} measure energy poverty in Greece using a combination of objective and subjective indicators, while \cite{Sareen2021} analyse how digital technologies are reshaping the governance of energy poverty. Recent work has further deepened the understanding of these issues. \cite{Bouzarovski2024} discuss how the energy transition can be made socially just by rethinking energy poverty policies in the context of decarbonisation. \cite{Berti2023} provide new evidence on the distributional impact of energy price increases in Italy, showing that income support must be paired with structural measures to be effective. \cite{Romero2023} quantify the potential of building retrofits to simultaneously alleviate energy poverty and reduce emissions, finding that targeted renovation of the worst-performing dwellings is the most cost‑effective strategy. At the European level, the Energy Poverty Advisory Hub (EPAH) released in 2024 a comprehensive set of indicators and policy recommendations that explicitly recognise the need to go beyond income‑based metrics \cite{EPAH2024}.

Despite these advances, most existing models treat energy as an ordinary consumption good and assume that any level of energy service can be purchased if income is high enough. This ignores a basic engineering fact: in extremely inefficient dwellings, the heat loss rate can exceed the heating system’s capacity, making it physically impossible to reach a minimum indoor temperature regardless of expenditure. A similar logic of non-convexities and poverty traps has been extensively studied in development economics since \cite{DasguptaRay1986} and \cite{BanerjeeNewman1993}, but has rarely been applied to energy poverty in advanced economies.

Our contribution is to fill this gap by building a microfounded model where energy poverty arises from a non-convex energy-to-comfort technology. The model generates a sharp feasibility threshold that separates structural from income-driven energy poverty, yields testable predictions about the regressive impact of energy price shocks, and provides a rigorous policy ranking: efficiency investments strictly dominate income transfers and price subsidies. Unlike earlier work, we explicitly derive the welfare consequences of the technological constraint and characterise the optimal policy mix. We also show graphically how a price increase expands the poverty region and how different policies counteract this expansion.

\section{A Model of Energy Services with a Feasibility Trap}
Consider a continuum of households indexed by \(i\). Each household has exogenous annual income \(y_i > 0\) and lives in a dwelling with energy efficiency \(\theta_i \in (0,1]\). Efficiency captures how much of each purchased unit of energy is converted into useful thermal comfort; it reflects insulation, window quality, and heating system performance. The market price of energy is \(p\).

Indoor thermal comfort \(h_i\) is produced by combining energy input \(x_i\) with the dwelling's efficiency according to a non-linear technology:
\begin{equation}
h_i = \theta_i x_i - \frac{\gamma}{2} x_i^2, \quad \gamma > 0.
\label{eq:prod}
\end{equation}
The concave shape captures physical diminishing returns: at high energy use, additional comfort gains become smaller because heat losses increase with the temperature difference between inside and outside. The parameter \(\gamma\) governs the speed of these diminishing returns.

A minimum comfort level \(\bar{h} > 0\) is required for healthy living (e.g., the WHO recommends a minimum indoor temperature). A household is in \emph{energy poverty} if \(h_i < \bar{h}\). The budget constraint is
\begin{equation}
c_i + p x_i = y_i,
\label{eq:budget}
\end{equation}
where \(c_i\) denotes consumption of all other goods. For simplicity we normalise the price of \(c_i\) to 1 and, importantly, we assume that no minimum consumption floor exists for non‑energy goods: a household can, if necessary, spend its entire income on energy. This delivers a clean upper‑bound income threshold. Introducing a subsistence requirement \(c_i \ge \underline{c} > 0\) would only tighten the budget constraint and amplify the structural trap described below, without altering its qualitative nature.

\begin{figure}[H]
\centering
\begin{tikzpicture}[scale=1.1]
    \draw[->] (0,0) -- (6.5,0) node[right] {Energy input $x$};
    \draw[->] (0,0) -- (0,5.2) node[above] {Comfort $h$};
    \draw[dashed, thick] (0,3) -- (6,3) node[right] {$\bar{h}$};
    \draw[domain=0:6, smooth, variable=\x, blue, thick] plot ({\x},{1.5*\x - 0.15*\x*\x});
    \node[blue] at (5.8,4.8) {$\theta_{\text{high}} > \theta^{\min}$};
    \draw[domain=0:6, smooth, variable=\x, red, thick] plot ({\x},{0.9*\x - 0.15*\x*\x});
    \node[red] at (5.8,1.8) {$\theta_{\text{low}} < \theta^{\min}$};
    \fill (2.76,3) circle (2pt) node[above right] {$x^*$};
\end{tikzpicture}
\caption{Comfort production for two different efficiency levels. For $\theta_{\text{high}}$ the curve crosses $\bar{h}$ at two points; the smallest feasible energy input is $x^*$. For $\theta_{\text{low}}$ the maximum comfort lies below $\bar{h}$, so the constraint can never be satisfied.}
\label{fig:production}
\end{figure}
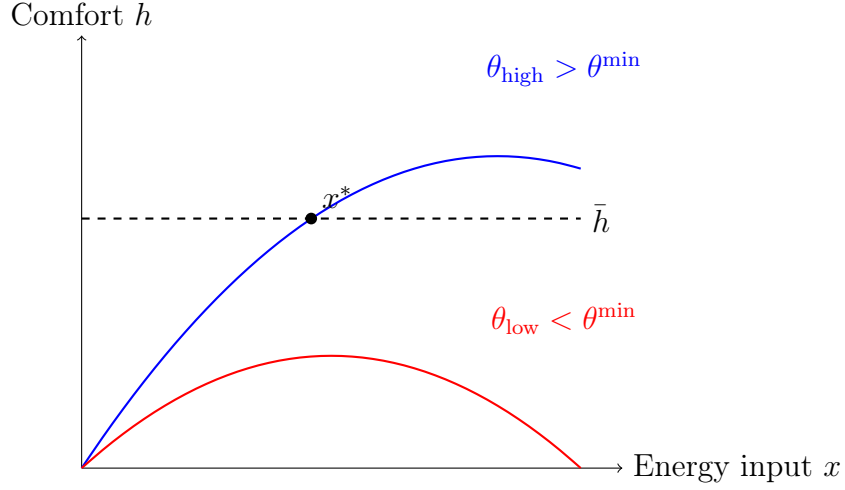

The central question is whether a household can physically reach \(\bar{h}\) given its dwelling. Setting \(h_i = \bar{h}\) in \eqref{eq:prod} yields the quadratic equation
\begin{equation}
\frac{\gamma}{2}x_i^2 - \theta_i x_i + \bar{h} = 0.
\label{eq:quad}
\end{equation}
Real solutions exist only if the discriminant is non-negative:
\begin{equation}
\theta_i^2 - 2\gamma \bar{h} \geq 0 \quad\Longrightarrow\quad
\theta_i \geq \sqrt{2\gamma\bar{h}} \equiv \theta^{\min}.
\label{eq:thr}
\end{equation}
Condition (\ref{eq:thr}) defines a \textbf{technological feasibility threshold}. If \(\theta_i < \theta^{\min}\), the quadratic has no real roots: the comfort function never reaches \(\bar{h}\) for any finite \(x_i\) (Figure~\ref{fig:production}). In other words, the dwelling is so inefficient that heat escapes faster than it can be supplied; throwing more energy at the problem only wastes resources without ever achieving minimum comfort.

\begin{figure}[H]
\centering
\begin{tikzpicture}[scale=1.2]
    \draw[->] (0,0) -- (6.2,0) node[right] {Efficiency $\theta_i$};
    \draw[->] (0,0) -- (0,5.5) node[above] {Income $y_i$};
    \draw[dashed, thick] (2.5,0) -- (2.5,5.2) node[above] {$\theta^{\min}$};
    \draw[domain=2.5:5.8, smooth, variable=\t, thick] plot ({\t}, {5* ( (\t - sqrt(\t*\t - 2*0.3*3))/0.3 ) });
    \fill[red, opacity=0.2] (0,0) rectangle (2.5,5.2);
    \node[red] at (1.25,2.5) {Structural};
    \fill[orange, opacity=0.2] (2.5,0) -- plot[domain=2.5:5.8] ({\x}, {5*( (\x - sqrt(\x*\x - 2*0.3*3))/0.3 )}) -- (5.8,0) -- cycle;
    \node[orange] at (4.2,1.5) {Income-driven};
    \fill[green, opacity=0.15] (2.5,5.2) -- plot[domain=5.8:2.5] ({\x}, {5*( (\x - sqrt(\x*\x - 2*0.3*3))/0.3 )}) -- (2.5,5.2);
    \node[green!50!black] at (4.5,4.5) {No poverty};
    \draw (0,0) -- (0,0);
\end{tikzpicture}
\caption{Regions of energy poverty in the efficiency–income space. The vertical dashed line marks the feasibility threshold $\theta^{\min}$. The curved line shows the minimum income needed to purchase the optimal energy input for a given efficiency. Structural poverty (left of the threshold) cannot be cured by income; income-driven poverty (below the curve) can.}
\label{fig:regions}
\end{figure}
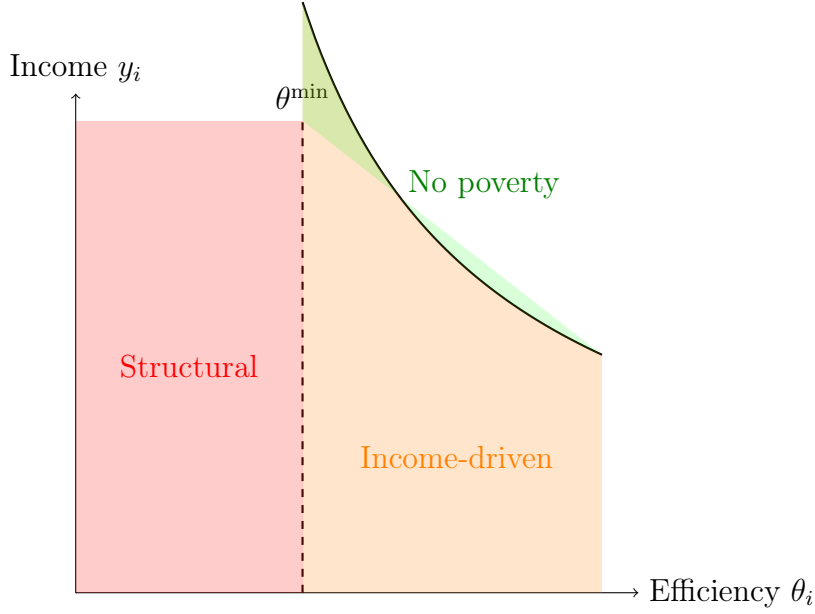

When \(\theta_i \geq \theta^{\min}\), the smallest energy input that delivers \(\bar{h}\) is the smaller root of \eqref{eq:quad}:
\begin{equation}
x^*(\theta_i) = \frac{\theta_i - \sqrt{\theta_i^2 - 2\gamma\bar{h}}}{\gamma}.
\label{eq:xstar}
\end{equation}
A household can afford this amount only if its income is high enough:
\begin{equation}
y_i \geq p\, x^*(\theta_i).
\label{eq:yth}
\end{equation}
Combining the physical and budget constraints yields three mutually exclusive regions in the \((\theta_i, y_i)\) space (Figure~\ref{fig:regions}):

\begin{itemize}
\item \textbf{Structural poverty:} \(\theta_i < \theta^{\min}\). Energy poverty is unavoidable regardless of income.
\item \textbf{Income-driven poverty:} \(\theta_i \geq \theta^{\min}\) and \(y_i < p\,x^*(\theta_i)\). The household could physically reach \(\bar{h}\) but cannot afford the necessary energy.
\item \textbf{No poverty:} \(\theta_i \geq \theta^{\min}\) and \(y_i \geq p\,x^*(\theta_i)\).
\end{itemize}

\section{The Regressive Impact of Energy Price Shocks}

\begin{figure}[H]
\centering
\begin{tikzpicture}[scale=1.1]
    \draw[->] (0,0) -- (6.5,0) node[right] {Efficiency $\theta_i$};
    \draw[->] (0,0) -- (0,8.5) node[above] {Income $y_i$};
    \draw[dashed, thick] (2.5,0) -- (2.5,8.2) node[above] {$\theta^{\min}$};
    \draw[domain=2.5:5.8, smooth, variable=\t, thick, blue] plot ({\t}, {4* ( (\t - sqrt(\t*\t - 2*0.3*3))/0.3 ) });
    \node[blue] at (4.5,3.5) {$y = p_0\,x^*(\theta)$};
    \draw[domain=2.5:5.8, smooth, variable=\t, thick, red] plot ({\t}, {6* ( (\t - sqrt(\t*\t - 2*0.3*3))/0.3 ) });
    \node[red] at (5.3,6.5) {$y = p_1\,x^*(\theta),\; p_1>p_0$};
    \fill[orange, opacity=0.35] 
        (2.5, {4* ( (2.5 - sqrt(2.5*2.5 - 1.8))/0.3 )}) --
        plot[domain=2.5:5.8] ({\x}, {6* ( (\x - sqrt(\x*\x - 1.8))/0.3 )}) --
        (5.8, {4* ( (5.8 - sqrt(5.8*5.8 - 1.8))/0.3 )}) --
        plot[domain=5.8:2.5] ({\x}, {4* ( (\x - sqrt(\x*\x - 1.8))/0.3 )}) -- cycle;
    \node[orange, font=\small\bf] at (3.8,4.2) {New poverty cases};
    \draw[->, orange, thick] (3.8,3.8) -- (3.2,2.5);
\end{tikzpicture}
\caption{Effect of a rise in the energy price from $p_0$ to $p_1$. The income threshold curve shifts upward, more steeply for low efficiency. The orange area shows households that fall into income‑driven poverty as a result of the shock.}
\label{fig:price_shock}
\end{figure}
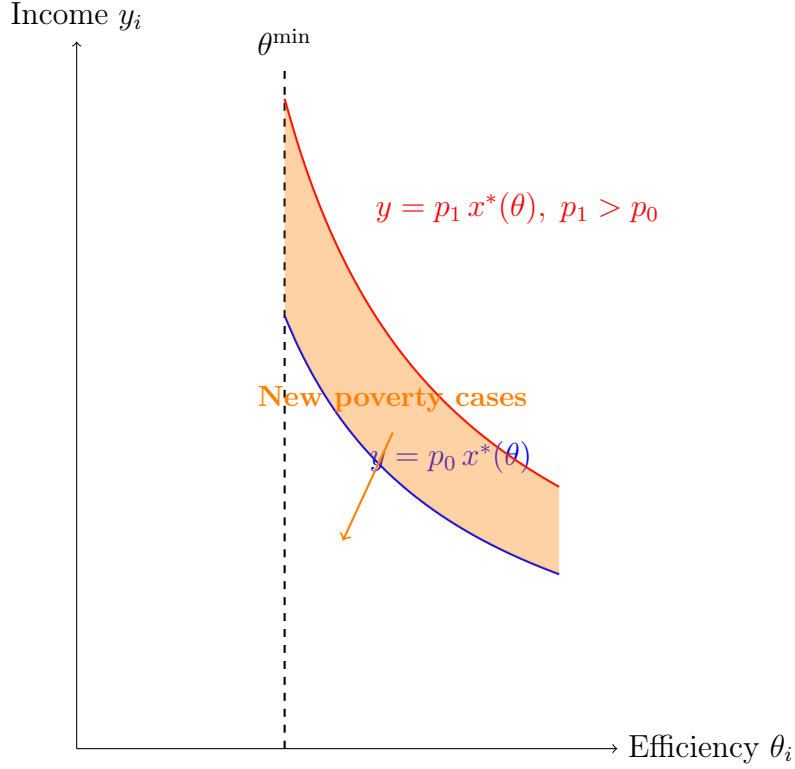

How does a rise in the energy price affect the poverty regions? For a given efficiency \(\theta_i \geq \theta^{\min}\), the required energy input \(x^*(\theta_i)\) is fixed by the thermal constraint. The income threshold therefore shifts in proportion to the price: \(y^{\text{threshold}}(\theta_i, p) = p\,x^*(\theta_i)\). An increase from \(p_0\) to \(p_1 > p_0\) rotates the threshold curve upward (Figure~\ref{fig:price_shock}). The shift is larger for low-efficiency dwellings because \(x^*(\theta_i)\) decreases with \(\theta_i\) (a better insulated home needs less energy to reach \(\bar{h}\)). Consequently, the regressive nature of energy price hikes is built into the technology: households in poorly insulated homes are not only more likely to be poor to begin with, but also suffer a proportionally larger increase in the required income to escape poverty.

Figure~\ref{fig:price_shock} makes the expansion of income-driven poverty visible. The orange shaded area between the old and new threshold curves represents households that were previously above the poverty line but fall into energy poverty after the price shock. The size of this area depends on the joint distribution of income and efficiency; it is largest when many households cluster just above the old threshold with relatively low efficiency.

\section{Policy Interventions and Their Mitigation Effects}

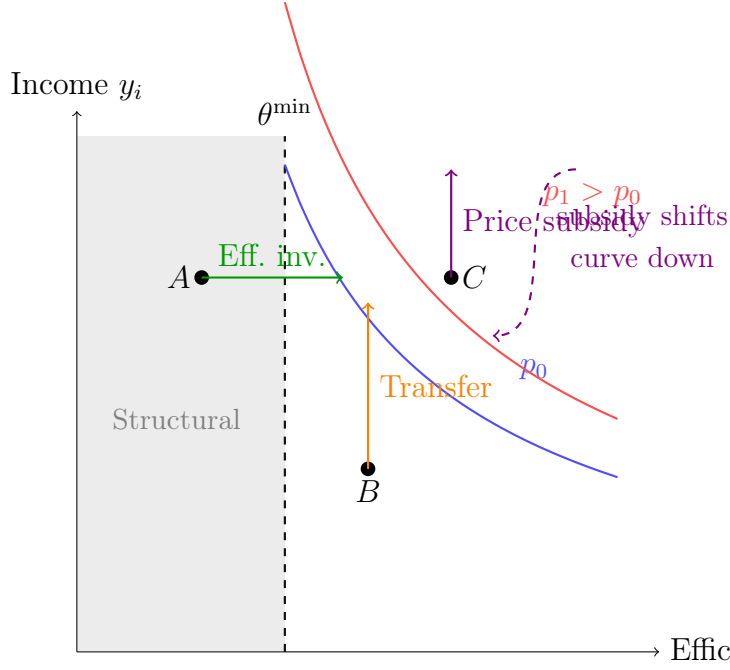
\begin{figure}[H]
\centering
\begin{tikzpicture}[scale=1.1]
    \draw[->] (0,0) -- (7,0) node[right] {Efficiency $\theta_i$};
    \draw[->] (0,0) -- (0,6.5) node[above] {Income $y_i$};
    \draw[dashed, thick] (2.5,0) -- (2.5,6.2) node[above] {$\theta^{\min}$};
    \draw[domain=2.5:6.5, smooth, variable=\t, thick, blue!70] plot ({\t}, {4.5* ( (\t - sqrt(\t*\t - 2*0.3*3))/0.3 ) });
    \node[blue!70] at (5.5,3.4) {$p_0$};
    \draw[domain=2.5:6.5, smooth, variable=\t, thick, red!70] plot ({\t}, {6.0* ( (\t - sqrt(\t*\t - 2*0.3*3))/0.3 ) });
    \node[red!70] at (6.2,5.5) {$p_1>p_0$};
    \fill[gray, opacity=0.15] (0,0) rectangle (2.5,6.2);
    \node[gray, font=\small] at (1.2,2.8) {Structural};

    \fill (1.5,4.5) circle (2.5pt) node[left] {$A$};
    \fill (3.5,2.2) circle (2.5pt) node[below] {$B$};
    \fill (4.5,4.5) circle (2.5pt) node[right] {$C$};

    \draw[->, thick, green!60!black] (1.5,4.5) -- (3.2,4.5) node[midway, above] {Eff.\ inv.};
    \draw[->, thick, orange] (3.5,2.2) -- (3.5,4.2) node[midway, right] {Transfer};
    \draw[->, thick, violet] (4.5,4.5) -- (4.5,5.8) node[midway, right] {Price subsidy};

    \draw[->, thick, violet, dashed] (6.0,5.8) to[out=180, in=0] (5.0,3.8);
    \node[violet, font=\small, align=center] at (6.8,5.0)
{subsidy shifts\\curve down};

\end{tikzpicture}
\caption{How different policies mitigate the effect of a price shock. Household $A$ (structural poverty) can only be helped by an efficiency investment (green arrow). Household $B$ (income‑driven poverty) can be lifted by a transfer (orange arrow). Household $C$ (just above the post‑shock threshold) could also benefit from a price subsidy that lowers the effective price (violet arrow). Only efficiency moves households across the $\theta^{\min}$ line, reducing structural poverty.}
\label{fig:policies}
\end{figure}

We now examine how three typical government policies can counteract the increase in energy poverty caused by a price shock. We assume the planner has a fixed budget \(B\) and can choose among a universal price subsidy, a lump-sum income transfer targeted to low-income households, and an efficiency investment program for the worst-performing dwellings. Figure~\ref{fig:policies} illustrates the mechanisms by showing how a concrete household in each region is affected.

\begin{itemize}
\item \textbf{Price subsidy} (reduction of \(p\) to \((1-\tau)p\)): The subsidy directly lowers the effective price, shifting the income threshold curve back downward (violet arrow). This helps households in the income‑driven zone, but does not alter the structural poverty region because \(\theta_i\) remains unchanged. Moreover, the subsidy benefits all consumers, including the non‑poor, generating fiscal waste.
\item \textbf{Income transfer} (a lump-sum payment \(T\) to all households below a certain income): The transfer moves households vertically upward (orange arrow). A family just below the income threshold can thus escape poverty. However, the transfer does nothing for those in structural poverty (\(\theta_i < \theta^{\min}\)), because their problem is not a lack of purchasing power but a physical impossibility.
\item \textbf{Efficiency investment} (an upgrade that raises \(\theta_i\) by \(\Delta\theta\) for the worst dwellings): This policy shifts households to the right (green arrow). Households originally left of \(\theta^{\min}\) can cross the threshold and leave structural poverty. Even those already feasible benefit, because a higher \(\theta\) reduces the required energy input \(x^*(\theta)\), lowering the income threshold and making comfort easier to afford.
\end{itemize}

Figure~\ref{fig:policies} summarises these effects with a stylised example. Three households are shown: one in structural poverty, one in income‑driven poverty, and one just above the post‑shock threshold. The arrows indicate how each policy moves them. The structural zone shrinks only under the efficiency policy.

\subsection{Formal Policy Ranking}

We now formalise the intuition presented above with two propositions.

\textbf{Proposition 1 (Regressive price effects).}  
Let \(\Delta EP\) denote the increase in the number of energy-poor households due to a price rise from \(p_0\) to \(p_1\). The marginal contribution of a household with efficiency \(\theta_i\) to \(\Delta EP\) is decreasing in \(\theta_i\). That is,
\[
\frac{\partial^2 EP}{\partial p\,\partial\theta_i} < 0.
\]
\textit{Proof.} For \(\theta_i \geq \theta^{\min}\), a household becomes poor if \(y_i < p\,x^*(\theta_i)\). The derivative of the income threshold with respect to price is \(x^*(\theta_i)\). Since \(x^*(\theta_i)\) is strictly decreasing in \(\theta_i\) (better efficiency reduces the necessary energy input), the threshold shift is larger for low \(\theta_i\). The cross-derivative is therefore negative. \hfill \(\square\)

\textbf{Proposition 2 (Efficiency dominance).}  
Suppose a fixed public budget can be spent either on a uniform price subsidy, a lump-sum transfer to the lowest-income households, or an efficiency upgrade for the least efficient dwellings. Then, if the economy has a positive mass of households with \(\theta_i < \theta^{\min}\), the reduction in the energy poverty headcount is strictly largest under the efficiency upgrade, followed by the transfer, and smallest under the price subsidy:
\[
\Delta EP_{\text{efficiency}} > \Delta EP_{\text{transfers}} > \Delta EP_{\text{subsidy}}.
\]
\textit{Proof sketch.} The price subsidy benefits all households proportionally to their energy consumption, which is highest among the non-poor, generating a large fiscal leakage. The transfer can be targeted, but it only moves households vertically; it cannot help those in structural poverty (\(\theta_i < \theta^{\min}\)). The efficiency investment shifts the \(\theta_i\) distribution rightward, which (i) moves some households across \(\theta^{\min}\), eliminating structural poverty for them, and (ii) lowers \(x^*(\theta_i)\) for those already feasible, reducing their income threshold. Near \(\theta^{\min}\), a small increase in \(\theta\) produces a discontinuity in welfare, while the same money given as income yields only a marginal utility gain. Hence the marginal poverty reduction per euro is highest for efficiency. \hfill \(\square\)

These propositions provide clear guidance for policy design. In the presence of a structural trap, a strategy that combines targeted efficiency investments with temporary income support for the most vulnerable is superior to any single-instrument approach.

\section{Conclusion}
This paper demonstrates that energy poverty can arise from a simple but overlooked physical constraint: below a certain dwelling efficiency, the required indoor comfort is physically unattainable. Our model produces a clear taxonomy of poverty types, explains the regressive impact of energy price shocks, and yields a sharp policy ranking. Efficiency investments dominate both income transfers and price subsidies because they shift the feasibility boundary itself, lifting households out of structural deprivation. The graphical analysis of the price shock and the countervailing policies provides an intuitive understanding of why well-insulated homes are not only an environmental necessity but also a social protection tool. The results urge a reorientation of anti-poverty strategies towards deep building renovation, particularly in countries with old and poorly insulated housing stocks. Such a strategy would simultaneously reduce energy poverty, lower carbon emissions, and shield vulnerable families from future energy price volatility.

\end{document}